\documentclass{WileyMSP-template}
\usepackage{CJK,multirow,graphics,fancyhdr,epstopdf}
\usepackage{amsmath,amsfonts,amssymb,bm,upgreek,mathrsfs,mathcomp,float}
\usepackage{adjustbox}

\begin{document}

\pagestyle{fancy}
\rhead{\includegraphics[width=2.5cm]{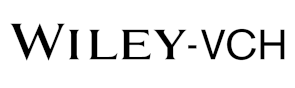}}

\title{Model of dark current in silicon-based barrier impurity band infrared detector devices}

\maketitle


\author{Mengyang Cui*}


\dedication{}

\begin{affiliations}

  Mengyang Cui\\
  Hangzhou Institute for Advanced Study, UCAS; Shanghai Institute of Technical Physics, Chinese Academy of Sciences; University of Chinese Academy of Sciences
  
  Email Address: KoitAgosti@gmail.com 
\end{affiliations}



\begin{abstract}

  Dark current in silicon-based blocked-impurity-band (BIB) infrared detectors has long been a critical limitation on device performance. This work proposes a chiral-phonon-assisted spin current model at 
  interfaces to explain the parabolic-like dark current behavior observed at low bias voltages. Concurrently, the Space Charge Limited Currents theory(SCLC) is employed to elucidate the dark current 
  generation mechanism across the entire operational voltage range.

\end{abstract}


\section{Introduction}
Silicon-based Blocked Impurity Band (BIB) detectors play a critically important role in astronomical detection\textsuperscript{[1,2]}. The photocurrent in these devices primarily 
originates from the photoexcitation of electrons from the impurity band into the conduction band within the absorbing layer\textsuperscript{[3,4]}.
While further increasing the doping concentration in the heavily doped absorbing layer can broaden the detector's spectral response range, 
excessively high doping levels can induce device breakdown. Furthermore, these devices are subject to significant dark currents, 
compounded by factors such as fabrication processes. This substantial noise severely compromises the detector's practical utility.
To provide a novel perspective, we propose for the first time a dark current model specifically for silicon-based BIB devices.
The silicon component of a Blocked Impurity Band (BIB) detector comprises distinct sections, including the blocking layer, 
interfaces, and bulk (absorbing) layer. This structure can be conceptually abstracted as a series connection of these principal 
components. While the electrical resistance and electric field distribution within each section are subject to variation with operating 
conditions such as temperature and applied voltage, the current density remains constant across any given transverse cross-section of 
the device. In silicon-based Blocked Impurity Band (BIB) detectors, the primary charge carriers are electrons injected at the cathode. 
Under standard operating conditions, these electrons traverse the absorbing region, pass through the interfacial layer(s), and are 
subsequently collected at the anode via the blocking layer.
Experimentally measured dark current characteristics typically exhibit distinct regimes: an initial nonlinear increase, 
followed by an abrupt transition to a linear ohmic conduction regime at a specific threshold voltage, culminating in a state of 
current saturation. Notably, current-voltage (I-V) characteristics of devices fabricated under non-ideal conditions occasionally 
manifest negative differential resistance (NDR). This observation suggests the presence of strongly localized electronic states within 
the device. Consequently, we postulate that a hopping mechanism between ionized donor sites dominates electron transport. 
Upon injection into the doped silicon material, electrons preferentially occupy available trap states. This trap-filling process 
governs conduction until the localized states are saturated. Subsequent carrier excitation into the extended states of the conduction 
band then facilitates a sudden, substantial increase in current, accounting for the observed abrupt rise.
Thus, the dark current mechanism reduces to analyzing the modulation of bulk current density within a fixed spatial domain as a function of applied bias (governing the electric field intensity) and temperature.

BIB devices operate at ultra-low temperatures where their fundamental physical properties remain incompletely characterized. Insights from superconducting experiments under analogous cryogenic conditions may provide critical guidance for elucidating their behavior.
Experimental evidence has demonstrated that both modified interfaces of silicon and doped bulk materials can achieve superconductivity\textsuperscript{[5,6]}. Chiral superconductivity in silicon-based materials was experimentally achieved, and
the superconducting transition temperature \(T_c\) is 
close to 10 K near the operating temperature regime of BIB devices. Doping and heterojunction engineering at silicon interfaces facilitate the formation of 
Cooper-like electron pairs, while phonon-assisted tunneling effects are rigorously validated through both experimental observations and theoretical modeling.
BIB device as a bulk material of silicon doped with phosphorus is considered to be a good candidate for this phenomenon.

In the gradient transition region of the bulk material, these temperature gradients may induce chiral phonons\textsuperscript{[7,8]}. Not originating from chiral molecular chains, these chiral phonons stem from electron transport between localized states and are only readily observable under low electric field conditions\textsuperscript{[9]}.
In the low-voltage regime, the dark current of BIB detectors initially displays a quadratic dependence on both voltage and temperature. As the electric field increases, the hopping capability of electrons between localized states is enhanced, resulting in greater complexity of electron transport pathways. Under such circumstances, the spin current becomes negligible. Depending on the distribution of dopant atoms in the material, phenomena including negative differential resistance and periodic current oscillations may arise\textsuperscript{[10]}. In highly ordered materials, the system typically transitions to a metallic state and eventually enters a saturation regime.


\section{Method}

Our theoretical framework posits that Cooper-like electron pairs\textsuperscript{[11]}—which do not strictly require antiparallel momenta—emerge from localized electronic states, in which the extended dimensions of the bulk material substantially exceed the coherence length of the wavefunction at low temperatures. Remarkably, even in systems dominated by Coulomb interactions, phonon-mediated attractive interactions facilitate the formation of Cooper pairs via the following mechanism: phonon-mediated attraction\textsuperscript{[12,13,14]}.
We estimate the average matrix element of local phonon-mediated attraction of the form
$$\overline{U_{\mathrm{e}-\mathrm{ph}}}=-\int d^3 \boldsymbol{r} d^3 \boldsymbol{r}^{\prime} 
g_{\mathrm{e}-\mathrm{ph}} \delta\left(\boldsymbol{r}-\boldsymbol{r}^{\prime}\right) 
\overline{\psi^2(\boldsymbol{r}) \psi^2\left(\boldsymbol{r}^{\prime}\right)}$$
Distances $r$ and $r^{'}$ are measured from the center of the local wavefunction, and the results 
can be obtained under a short-distance cutoff on the order of the lattice constant
$$
\overline{U_{\mathrm{e}-\mathrm{ph}}} \approx-\frac{\pi}{\mathcal{I}_{3-d_2}} \frac{g_{\mathrm{e}-\mathrm{ph}}}{a^3}\left(\frac{a}{\xi_{\mathrm{loc}}}\right)^{d_2}
$$
\medskip
The Coulomb repulsion between two electrons is
$$
\overline{U_C}=\int d^3 \boldsymbol{r} d^3 \boldsymbol{r}^{\prime} 
\frac{e^2}{\varepsilon_1\left|\boldsymbol{r}-\boldsymbol{r}^{\prime}\right|} 
\overline{\psi^2(\boldsymbol{r}) \psi^2\left(\boldsymbol{r}^{\prime}\right)}=
\frac{\mathcal{I}_{4-d_2}}{\mathcal{I}_{3-d_2}} \frac{e^2}{\varepsilon_1 \xi_{\mathrm{loc}}}
$$
The ratio of the phonon-mediated attraction to the Coulomb repulsion between two electrons is
$$
\frac{\left|\overline{U_{\mathrm{e}-\mathrm{ph}}}\right|}{\overline{U_C}} \approx \lambda_0 \frac{\pi}{\mathcal{I}_{4-d_2}}\left(\frac{a}{\xi_{\mathrm{loc}}}\right)^{d_2-1} \frac{\varepsilon_1}{e^2 v_0 a^2} \approx 2 \lambda_0
$$
Substituting the existing numerical results into the model yields the ratio $> 1$\textsuperscript{[15,16]}, 
indicating that phonon-mediated electrons pair formation is energetically favorable\textsuperscript{[17,18,19]}.
At the interface where the concentration of doped atoms rises sharply, electrons hop among various 
localized states. Due to the existence of the concentration gradient and the assistance of 
phonons with non-zero angular momentum\textsuperscript{[20]} for their tunneling, the interaction with the spin bath is
$$
\begin{aligned}
V= & \sum_{k, k^{\prime}} \sum_j e^{i\left(k^{\prime}-k\right) X_j}\left[J_{k \uparrow, k^{\prime} \uparrow}^j S_z^j c_{k \uparrow}^{\dagger} c_{k^{\prime} \uparrow}-J_{k \downarrow, k^{\prime} \downarrow}^j S_z^j c_{k \downarrow}^{\dagger} c_{k^{\prime} \downarrow}\right. \\
& \left.+J_{k \uparrow, k^{\prime} \downarrow}^j S_{-}^j c_{k \uparrow}^{\dagger} c_{k^{\prime} \downarrow}+J_{k \downarrow, k^{\prime} \uparrow}^j S_{+}^j c_{k \downarrow}^{\dagger} c_{k^{\prime} \uparrow}\right]
\end{aligned}
$$
here $J_{k \sigma, k^{\prime} \sigma^{\prime}}^j$ is the position and wave-vector-dependent coupling.
also the observation that resistivity increases with decreasing temperature at low temperatures (Kondo effect)\textsuperscript{[21]} indicates that the BIB device resides in a state of short-range magnetic ordering, characterized by relatively high spin polarizability, the overall macroscopic current turns out 
to be a helical current\textsuperscript{[22,23]} with a continuously changing radius. This kind of current leads to the 
existence of a magnetic field, that is, an equivalent Zeeman field\textsuperscript{[24,25,26,27,28]}.
$$
B_{\mathrm{eff}}=\eta \mathscr{E} =\eta_0 e^{-\frac{T}{T_c}} \mathscr{E}
$$
Here, $\eta_0$ represents a constant of the device, and \(T_c\) is the critical transition temperature of preformed cooper pairs.
Under the influence of an external electric field $\mathscr{E}$, the Zeeman energy-splitting term, as we 
mentioned above, becomes a crucial component for each spin within the Bogoliubov-de Gennes (BdG) 
Hamiltonian. This term introduces a splitting of energy levels due to the interaction between the 
magnetic moment of the spins and the applied electric field, leading to distinct energy states that 
are essential for understanding the quantum mechanical behavior of superconductors and other complex materials.

According to the Bogoliubov-de Gennes (BdG)\textsuperscript{[29]} equations, the Hamiltonian for the part  with significant doping concentration gradients is
 constructed as $H=\frac{1}{2} \sum_{\boldsymbol{k}} \Psi_{\boldsymbol{k}}^{\dagger} \mathscr{H}(\boldsymbol{k}) \Psi_{\boldsymbol{k}}$, 
 $\Psi_{\boldsymbol{k}}^{\dagger}=\left(c_{\boldsymbol{k}, \uparrow}^{\dagger}, c_{\boldsymbol{k}, \downarrow}^{\dagger}, c_{-\boldsymbol{k}, \uparrow}, c_{-\boldsymbol{k}, \downarrow}\right)$
 where $c_{\boldsymbol{k}, \sigma }^{\dagger}$ represents the creation (annihilation) operator of the electron
 with spin $\sigma$.
The Zeeman energy level splitting effect of \( \mathscr{E} \) under an applied electric field is included in the following equation
$$
\mathscr{H}(\boldsymbol{k})=\left(\begin{array}{cccc}
\xi_{\boldsymbol{k}}+\eta \mathscr{E} & 0 & 0 & \Delta \\
0 & \xi_{\boldsymbol{k}}-\eta \mathscr{E} & -\Delta & 0 \\
0 & -\Delta & -\xi_{-\boldsymbol{k}}-\eta \mathscr{E} & 0 \\
\Delta & 0 & 0 & -\xi_{-\boldsymbol{k}}+\eta \mathscr{E}
\end{array}\right)
$$
where $\Delta$ represents the binding energy of electrons at the same lattice site, and $\xi_{\boldsymbol{k}}=\frac{\hbar^2k^2}{2m}-\mu $
with $\mu$ denoting the chemical potential.
Under this model, the expression for the spin Bogoliubov quasiparticles current density can be obtained according to Reference \textsuperscript{[30]}.
$$
j_s=-\frac{e^2 \tau_{\mathrm{n}} \sqrt{2 m} \mathscr{E}}{4 \pi^2 \hbar^3 k_B T} \int_{-\mu}^{\xi_c} d \xi \frac{\xi}{\sqrt{\xi^2+\Delta^2}}(\xi+\mu)^{\frac{3}{2}}\left\{\cosh ^{-2}\left(\frac{\sqrt{\xi^2+\Delta^2}+\eta \mathscr{E}}{2 k_B T}\right)-\cosh ^{-2}\left(\frac{\sqrt{\xi^2+\Delta^2}-\eta \mathscr{E}}{2 k_B T}\right)\right\}
$$
\( \tau_n \) is the relaxation time in the normal state, and the relationship between it and the relaxation 
time in the pseudo-superconducting state is \( \tau_n = \frac{|\xi_{\mathrm{k}} | }{\sqrt{\xi_{\mathrm{k}^2}+\Delta^2}} \).
The spin current carried by quasiparticles flows along the screw axis and exhibits a quadratic 
dependence on the electric field in the low-field range. 
\begin{figure}[H]
  \includegraphics[width=\linewidth]{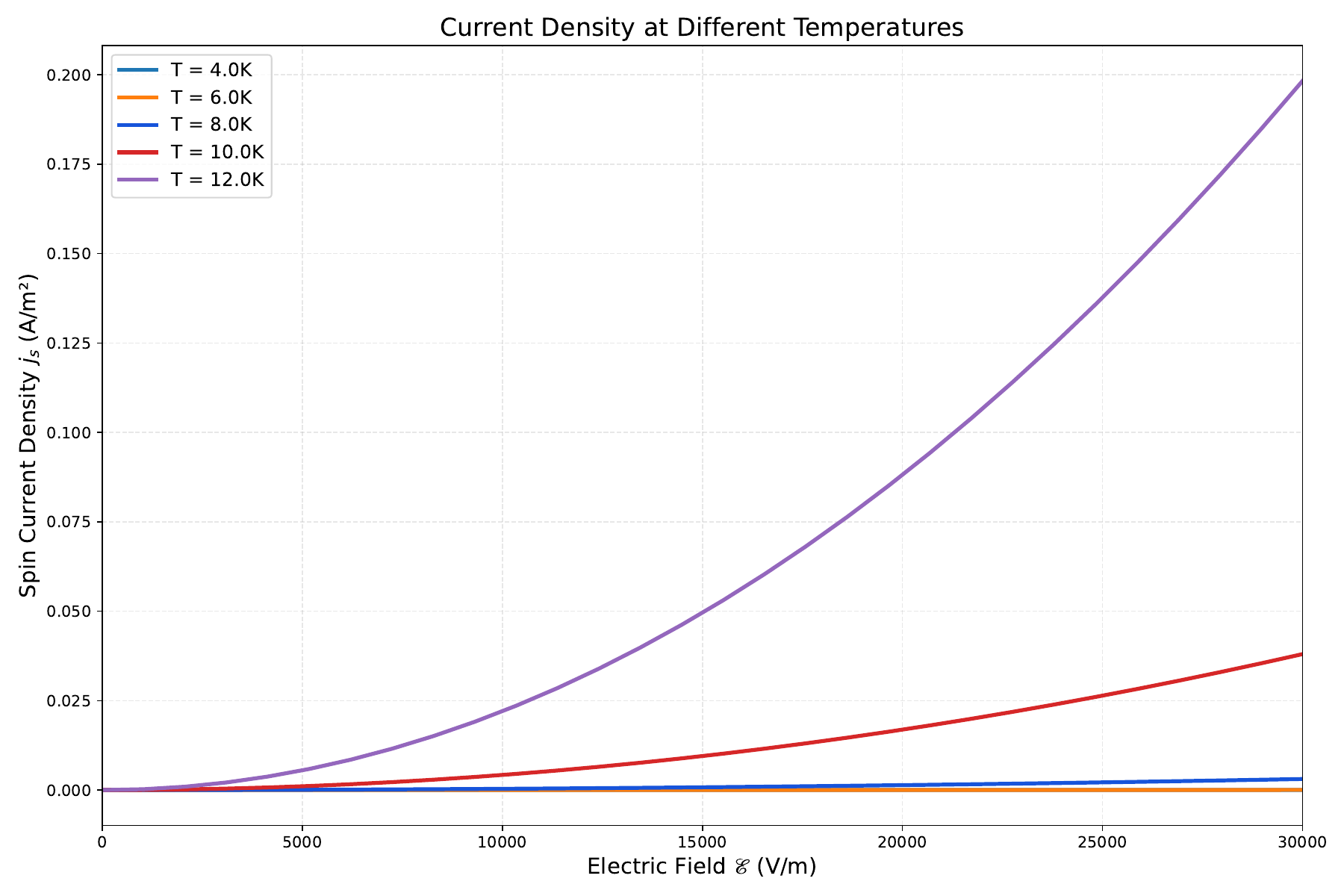}
  \caption{Theoretical prediction curves of spin current density at the absorber/blocking-layer interface. Key parameters include: relaxation time of 10 ps, chemical potential at 0.15 eV, cutoff energy of 0.05 eV, transition temperature of 60 K, and chiral phonon coupling coefficient of 1e-15 $eV \cdot  m \cdot V^{-1}$.}
  \label{fig:boat1}
\end{figure}


Excessively high electric field strength and excessively large carrier concentration will both disrupt the current density model that depends on the existence of localization for its functionality. Based on experimental measurement data, we conclude that this model typically ceases to be valid within the applied voltage range exceeding approximately 0.2 V.
Since the doping concentration of the blocking layer is several orders of magnitude lower than that of the absorption layer, the applied bias voltage basically drops across the blocking layer. For the simplicity of the model, we assume that the entire applied bias voltage falls across the blocking layer. As the absorption layer has a high doping concentration and is not the limiting factor for current, the interface between the two regions loses a significant hindering effect on carrier transport in the high-voltage range. Thus, discussing the scenario where the entire voltage is applied to the blocking layer is essentially equivalent to the overall current behavior.
We consider that the carriers in this segment (about 5$\mu m$) must fix Possion equation:
$\frac{d F}{d x}=\frac{e}{\epsilon}\left(\rho+n_t(\rho)-N_D\right)$, and
we assume that local thermal equilibrium is established within this region. When the 
transition region is excessively narrow and the doping concentration gradient is too abrupt, current flow becomes severely limited by the interfacial 
potential barrier. Electrons traversing this interface from the absorption layer to the blocking layer encounter greater difficulty compared to transport in 
the reverse direction. This asymmetry arises because electron conduction primarily occurs via hopping between donor ions. Electrons originating from regions of 
lower doping concentration transition more readily into higher-concentration regions. We attribute this phenomenon to the observed lower forward-bias current relative 
to reverse-bias current in certain devices. However, making the transition region less abrupt can improve the symmetry of current flow under both forward and reverse bias conditions.

We propose that, statistically, most dopant electrons preferentially occupy trap states initially. The initial Ohmic rise of the current is dominated by electrons being thermally excited and released from deep trap states, as well as driven by the external field, to become free carriers. Meanwhile, due to variations in the intensity of interfacial chiral phonons caused by process-related factors, the current in the low-voltage range is approximately linear. Subsequently, as the applied voltage increases, when deep traps are filled with carriers, the role of dopant energy levels gradually becomes prominent. Combined with the Frenkel effect (where the electric field reduces the ionization energy), more carriers are released from dopant states or trap states, participate in transport, and drive the current into the power-law rise region.
The $N_D$ is the density of ionized donors and the $n_t$ is the density of electrons in the traps which rely on the density of  free electrons
and the areal density of current is defined by\textsuperscript{[31]}

$$J_n=\mu_n(e n \mathcal{E}+k T \nabla n)+e \mu_t n_{t D} \mathcal{E}$$ , the third term is the tunneling current part, where the $n_{tD} = N_t f(E_t) \left[1 - f(E_D)\right]$ is the 
density of tunneled electrons and the $\mu_t$ is the mobility of the tunneling electrons.
$$
\mu_t = \frac{e d}{4 k T} \nu \exp\left( \frac{E_t - E_D}{k T} \right)
$$
The $n_t$ is defined by
$$
\begin{aligned}
n_t & =\frac{g_1 N_t n N_c \exp \left(-E_1 / k T\right)}{N_c^2+g_1 n N_c \exp \left(-E_1 / k T\right)+g_2 n^2 \exp \left(-E_2 / k T\right)} \\
& +\frac{2 g_2 N_t n^2 \exp \left(-E_2 / k T\right)}{N_c^2+g_1 n N_c \exp \left(-E_1 / k T\right)+g_2 n^2 \exp \left(-E_2 / k T\right)}
\end{aligned}
$$
where $E_1=E_t-E_c+\delta E^{F r}+\delta E_1^{S c r}$ ,$E_2=2\left(E_t-E_c\right)+U+\delta E^{F r}+\delta E_2^{S c r}$,
$E_1$ is the effective trap level for single occupancy and $E_2$ is for double occupancy, $g_1$ and $g_2$ are degeneracy factors, $N_c$ is the effective conduction band DOS. 
Electrons are dominated by single occupancy in the low-voltage region, while beyond approximately 1 V, the occurrence of double occupancy in some traps leads to a relatively slower current increase. For devices fabricated using materials not grown via molecular beam epitaxy (MBE) in the early stage, the presence of a relatively flat current growth region within a certain range beyond 1 V bias voltage may be attributed to the high density of deep-level traps.
The BIB device was theoretically modeled as a cubic structure so that we can constructed an areal current density model incorporating key 
phenomena including the Frenkel effect\textsuperscript{[32]}, single/double electron occupancy at lattice sites, and other relevant physical properties. 
Material-specific parameters were then applied to this model to generate the predicted results shown in and Fig. 3.
\begin{figure}[H]
\begin{adjustbox}{width=\linewidth}
\includegraphics{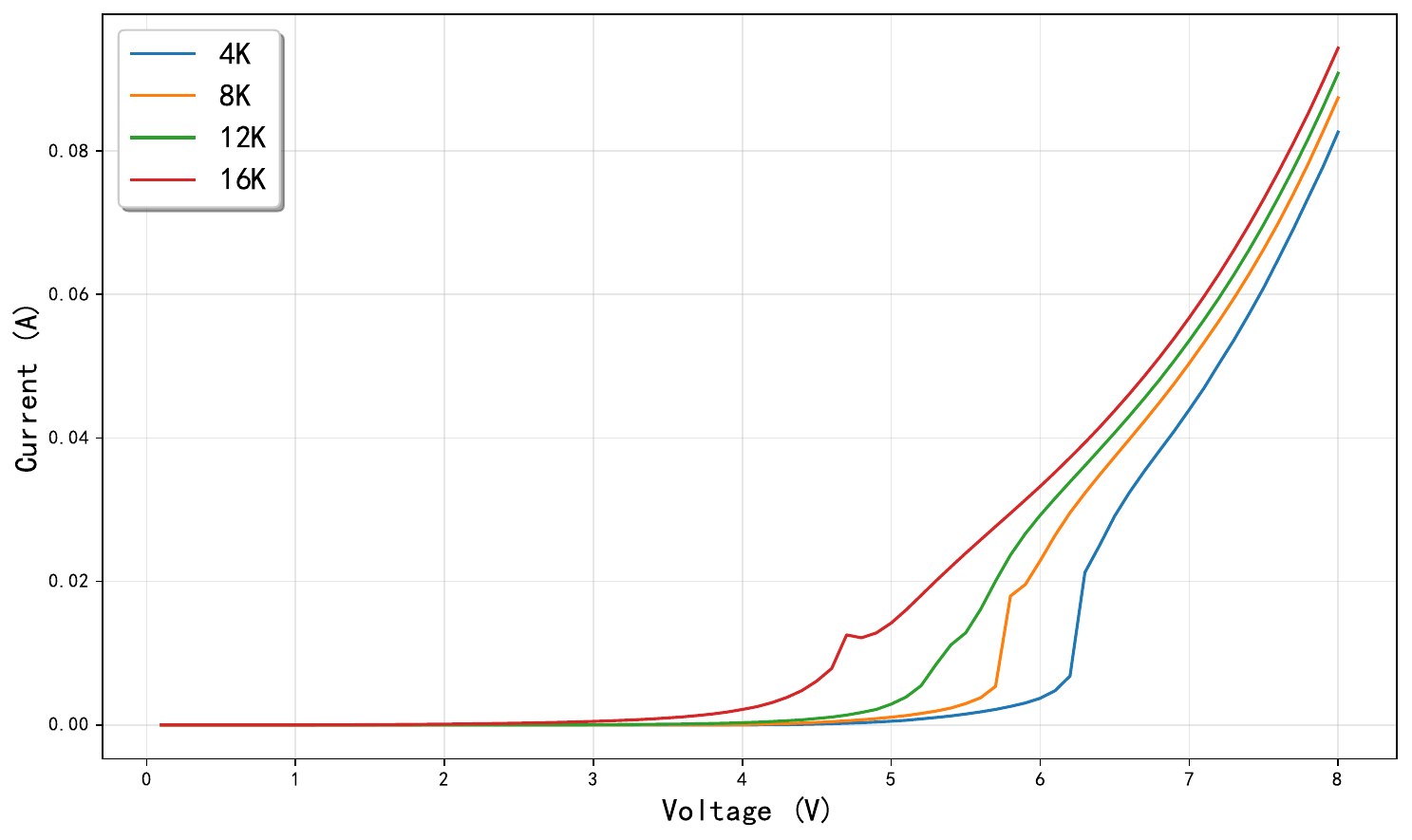}
\end{adjustbox}
\caption{\label{picture_2}Dark Current-Voltage Characteristics of Si:P BIB Devices at Different Temperatures. We generally observe a gradual variation in the curvature of the curves; the curvature remains nearly constant in the high-bias region, manifesting an insulator-to-metal transition (IMT) driven by the applied bias.}
\end{figure}

From the typical dark current measurement results of the device presented in Figure 2, it can be observed that the high-voltage region exhibits metallic characteristics with constant conductivity, whereas the dark current in the low-voltage region requires simultaneous increases in temperature and voltage to rise significantly. This behavior in the low-voltage regime is attributed to carrier trapping within the device. Only when elevated temperature and voltage enhance the carrier concentration sufficiently to fill the trap levels does the current start to increase remarkably beyond approximately 4 V, followed by a transition to the metallic state.
We note that the absorption region of the BIB device features a relatively high doping concentration. Some devices even directly incorporate a segment with a doping concentration on the order of $10^{18} cm^{-3}$ as the collector. Under conditions such as trap levels and low-temperature-induced localization, various factors that enhance electric field screening coexist. Consequently, the I-V characteristics in the high-voltage region display more pronounced metallic traits. We therefore propose that the overall dielectric constant of the device under DC voltage is significantly enhanced by the heavily doped regions, deviating from the intrinsic dielectric constant of bulk silicon (11.7).
A relatively large equivalent dielectric constant of 60,000 was adopted in the calculations presented in Figure 3. The calculated results at a device temperature of 4 K show good agreement with the experimental measurements. For the higher temperature range, the trends of the measured values are qualitatively consistent. This is because increasing temperature generally causes the curves to shift to the upper left. The disappearance of the current jump phenomenon around 3.5 V can be attributed to the increased carrier concentration induced by higher temperatures, which facilitates the filling of trap levels. Consequently, the jump in the current slope observed around 3.5 V at 4 K transforms into an abrupt increase in current density near 0.7 V at 16 K.
\begin{figure}[H]
\begin{adjustbox}{width=\linewidth}
\includegraphics{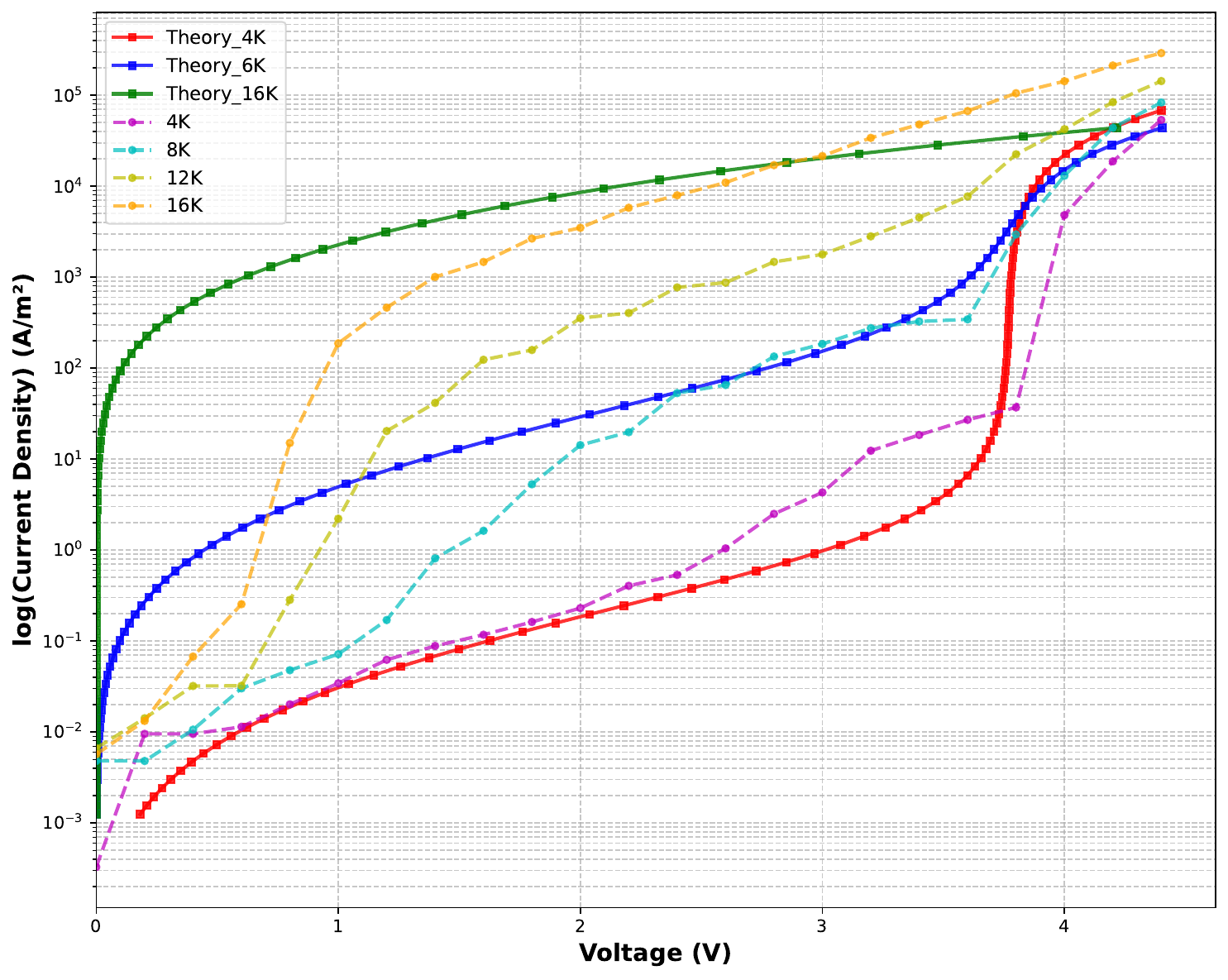}
\end{adjustbox}
\caption{\label{picture_3} Theoretical predictions of areal current density profiles derived from space-constrained charge transport theory. 
Calculations determine the required voltages to achieve target areal current densities at varying temperatures, utilizing the following key parameters: 
relative permittivity of 60000, on-site potential is 0.52eV, effective electron mass of 0.20 $m_e$, dopant energy level at -0.045eV, trap density of 1e24$m^{-3}$, and doping concentration of 1e19$m^{-3}$ }
\end{figure}
The Space-Charge-Limited Current (SCLC) theory exhibits limited applicability in the voltage range below 1 V. This is because it fails to account for physical mechanisms such as localization and carrier freezing, and assumes that the dominant carriers are purely electrons. The SCLC theory is more suitable for describing the transport properties of relatively ideal homogeneous materials under applied voltages, rather than inhomogeneous devices. However, when the carrier concentration in the BIB device is sufficiently high, it can be approximately regarded as a homogeneous material. Although the predictions in Figure 3 show deviations in the current magnitude over certain voltage ranges, the trend of current variation is generally correct. The main source of error is likely attributed to the fact that the dynamic variations of physical parameters in the calculation code are not precisely known.

\begin{figure}[H]
  \begin{adjustbox}{width=\linewidth}
  \includegraphics{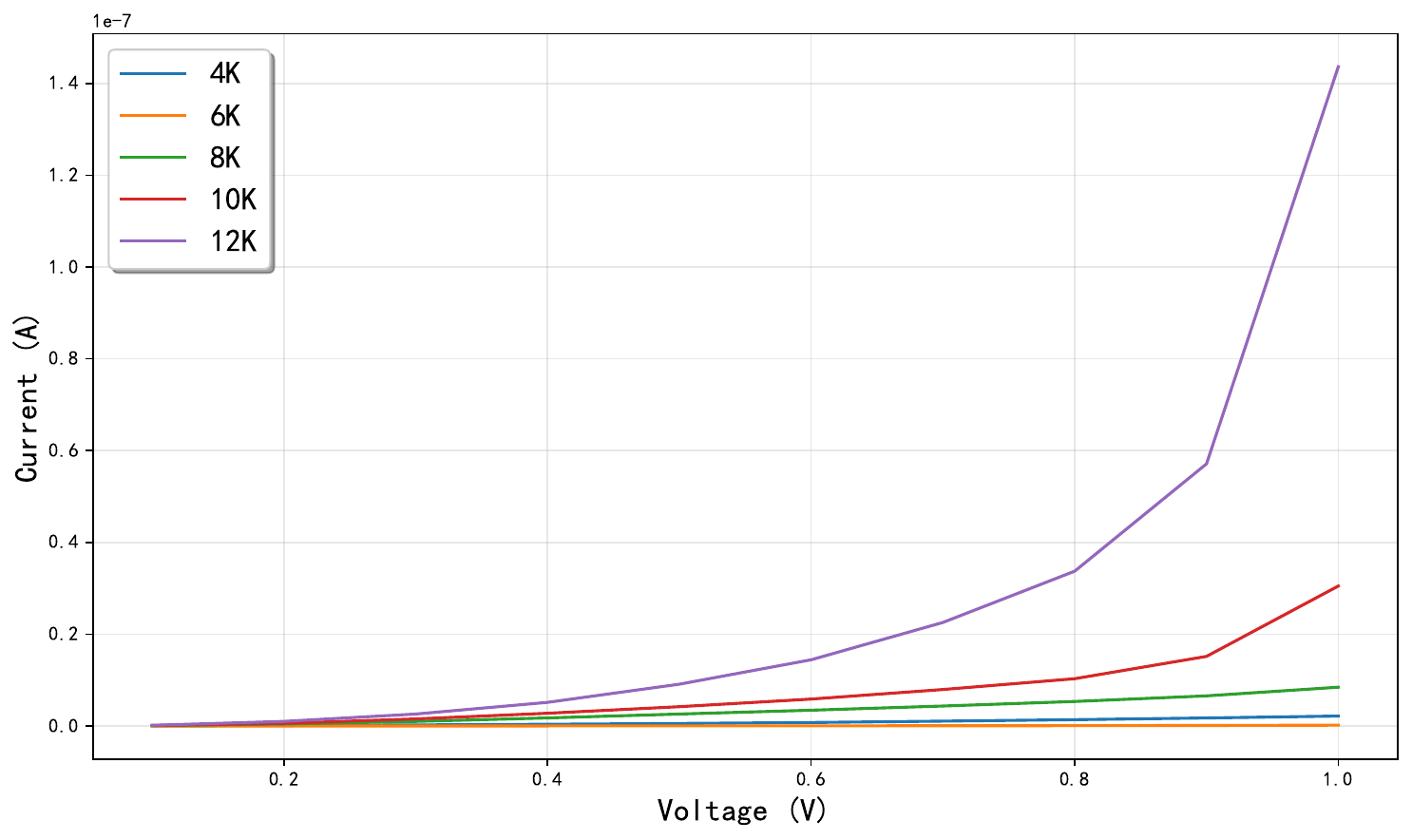}
  \end{adjustbox}
  \caption{\label{picture_4}Typical dark current curves of a Si:P BIB device in the low-voltage regime. It can be observed that the curves are approximately parabolic in shape, with the current reaching its minimum at 6 K. This phenomenon suggests that the formation efficiency of unstable Cooper pairs at the device interface is the highest at 6 K, which is consistent with our theoretical analysis.}
  \end{figure}
\section{Conclusion}
We propose an interfacial chiral phonon-assisted charge transport theory to model dark current transport in Blocked Impurity Band (BIB) devices under low-bias 
regimes. The quasi-helical transport image at the absorber/blocking-layer interface explains the asymmetric dark current characteristics under opposite bias 
polarities, attributable to electron cloud localization: electron transport from high-doping to low-doping regions exhibits lower energy barriers. Furthermore, 
based on space-constrained charge transport theory, we propose a sequential trapping-detrapping mechanism where electrons first occupy trap states before 
thermal excitation to the conduction band. This model quantitatively predicts dark current magnitudes, offering theoretical guidance for suppressing dark 
current in BIB detectors. 




\medskip
\textbf{Acknowledgements} \par 
This work is supported by the Research Funds of Hangzhou Institute for Advanced Study, UCAS.

\medskip

%

\textbf{References}\\

1 Xiao, Y., Zhu, H., Deng, K. et al. Progress and challenges in blocked impurity band infrared detectors for space-based astronomy. Sci. China Phys. Mech. Astron. 65, 287301 (2022). https://doi.org/10.1007/s11433-022-1906-y

2 Liu, KM., Hsieh, YE. The effects of source doping concentration and doping gradient on the ON-state current of Si nanowire TFETs. J Comput Electron 22, 209\textendash218 (2023). https://doi.org/10.1007/s10825-022-01995-6 

3 Frank L. Madarasz, Frank Szmulowicz; Barrier formation in graded Hg$_{1-x}$Cd$_{x}$Te heterojunctions. J. Appl. Phys. 15 October 1987; 62 (8): 3267\textendash3277. https://doi.org/10.1063/1.339333

4 Poduval, Prathyush P. and Das Sarma, Sankar. Anderson localization in doped semiconductors. Phys. Rev. B 107, 174204 (2023). https://link.aps.org/doi/10.1103/PhysRevB.107.174204

5 Ming, F., Wu, X., Chen, C. et al. Evidence for chiral superconductivity on a silicon surface. Nat. Phys. 19, 500–506 (2023). https://doi.org/10.1038/s41567-022-01889-1  

6 Bustarret, E., Marcenat, C., Achatz, P. et al. Superconductivity in doped cubic silicon. Nature 444, 465–468 (2006). https://doi.org/10.1038/nature05340

7 Romao, Carl P. Anomalous thermal expansion and chiral phonons in ${\mathrm{BiB}}_{3}{\mathrm{O}}_{6}$. Phys. Rev. B 100, 060302(R) (2019). https://link.aps.org/doi/10.1103/PhysRevB.100.060302

8 Hamada, Masato and Minamitani, Emi and Hirayama, Motoaki and Murakami, Shuichi. 
Phonon Angular Momentum Induced by the Temperature Gradient. Phys. Rev. Lett. 121, 175301 (2018). 

https://doi.org/10.1103/PhysRevLett.121.175301

9 Wang, Mao and Debernardi, A. and Berenc\'en, et al. Breaking the Doping Limit in Silicon by Deep Impurities. Phys. Rev. Applied 11, 054039 (2019). 
https://doi.org/10.1103/PhysRevApplied.11.054039

10  D. I. Aladashvili, Z. A. Adamiya, K. G. Lavdovskii, et al., Pis’ma Zh. Éksp. Teor. Fiz. 47, 390 (1988) [JETP Lett. 47, 466 (1988)]

11 Cliff Chen et al. ,Signatures of a spin-active interface and a locally enhanced Zeeman field in a superconductor-chiral material heterostructure.Sci. Adv.10,eado4875(2024). https://doi.org/10.1126/sciadv.ado4875 

12 B. I. Shklovskii; Half-century of Efros–Shklovskii Coulomb gap: Romance with Coulomb interaction and disorder. Low Temp. Phys. 1 December 2024; 50 (12): 1101–1112. https://doi.org/10.1063/10.0034343

13 Wu, X., Li, X., Kang, W. et al. Topology-induced chiral photon emission from a large-scale meron lattice. Nat Electron 6, 516–524 (2023). https://doi.org/10.1038/s41928-023-00990-4

14 Kim, Minho and Timmel, Abigail and Ju, Long and Wen, Xiao-Gang. Topological chiral superconductivity beyond pairing in a Fermi liquid. Phys. Rev. B 111, 014508 (2025). 

https://doi.org/10.1103/PhysRevB.111.014508

15 Lindinger, Jakob and Buchleitner, Andreas and Rodr\'{\i}guez, Alberto. Many-Body Multifractality throughout Bosonic Superfluid and Mott Insulator Phases. Phys. Rev. Lett. 122, 106603 (2019). 

https://doi.org/10.1103/PhysRevLett.122.106603

16 Galeazzi, M. and Liu, D. and McCammon, D. and Rocks, L. E. and Sanders, W. T. and Smith, B. and Tan, P. and Vaillancourt, J. E. and Boyce, K. R. and Brekosky, R. P. and Gygax, J. D. and Kelley, R. L. and Kilbourne, C. A. and Porter, F. S. and Stahle, C. M. and Szymkowiak, A. E.
Hot-electron effects in strongly localized doped silicon at low temperature. Phys. Rev. B 76, 155207 (2007). 

https://link.aps.org/doi/10.1103/PhysRevB.76.155207

17 Igor Poboiko and Mikhail Feigel'man. Mean-field theory of first-order quantum superconductor-insulator transition. SciPost Phys. 17, 066 (2024). https://scipost.org/10.21468/SciPostPhys.17.2.066

18 Zhang, Lifa and Niu, Qian. Angular Momentum of Phonons and the Einstein--de Haas Effect. PhysRevLett.112.085503 (2014). https://doi.org/10.1103/PhysRevLett.112.085503

19 Seung Gyo Jeong et al. ,Unconventional interlayer exchange coupling via chiral phonons in synthetic magnetic oxide heterostructures.Sci. Adv.8,eabm4005(2022). https://doi.org/10.1126/sciadv.abm4005

20 Li, Xiaozhe and Xia, Chongqun and Pan, Yang and Gao, Mengnan and Chen, Hao and Zhang, Lifa. Topological chiral phonons along the line defect of intralayer heterojunctions. Phys. Rev. B 104, 054103 (2021). https://doi.org/10.1103/PhysRevB.104.054103

21 Im, H., Lee, D.U., Jo, Y. et al. Observation of Kondo condensation in a degenerately doped silicon metal. Nat. Phys. 19, 676–681 (2023). https://doi.org/10.1038/s41567-022-01930-3

22 Sun, H., Huang, Y., He, M. et al. Chiral and helical states in selective-area epitaxial heterostructure. Commun Phys 6, 204 (2023). https://doi.org/10.1038/s42005-023-01328-4

23 Maeda, Kazuki and Fukaya, Yuri and Yada, Keiji and Lu, Bo and Tanaka, Yukio and Cayao, Jorge. Classification of pair symmetries in superconductors with unconventional magnetism. PhysRevB.111.144508 (2025). https://doi.org/10.1103/PhysRevB.111.144508

24 Xiong, Guohuan and Chen, Hao and Ma, Dengke and Zhang, Lifa. Effective magnetic fields induced by chiral phonons. Phys. Rev. B 106, 144302 (2022). https://doi.org/10.1103/PhysRevB.106.144302

25 Lunde, Anders Mathias and Platero, Gloria. Helical edge states coupled to a spin bath: Current-induced magnetization. Phys. Rev. B 86, 035112 (2012). https://doi.org/10.1103/PhysRevB.86.035112

26 Fukaya, Yuri and Hashimoto, Tatsuki and Sato, Masatoshi and Tanaka, Yukio and Yada, Keiji. Spin susceptibility for orbital-singlet Cooper pair in the three-dimensional ${\mathrm{Sr}}_{2}{\mathrm{RuO}}_{4}$ superconductor. Phys. Rev. Research 4, 013135 (2022). https://doi.org/10.1103/PhysRevResearch.4.013135

27 Felix G. G. Hernandez et al. ,Observation of interplay between phonon chirality and electronic band topology.Sci. Adv.9,eadj4074(2023).https://doi.org/10.1126/sciadv.adj4074

28 Hao Chen, Weikang Wu, Kangtai Sun, Shengyuan A. Yang, Lifa Zhang; Electrically controllable chiral phonons in ferroelectric materials. Appl. Phys. Lett. 26 February 2024; 124 (9): 092201. 

https://doi.org/10.1063/5.0196731

29 Romano, Alfonso and Noce, Canio and Cuoco, Mario. Magnetic exchange interaction in a spin valve with a chiral spin-triplet superconductor. Phys. Rev. B 109, 134516 (2024). 

https://doi.org/10.1103/PhysRevB.109.134516

30 Dapeng Yao, Mamoru Matsuo, Takehito Yokoyama; Electric field-induced nonreciprocal spin current due to chiral phonons in chiral-structure superconductors. Appl. Phys. Lett. 15 April 2024; 124 (16): 162603. https://doi.org/10.1063/5.0207915

31 Zhang, X.-G. and Pantelides, Sokrates T. Theory of Space Charge Limited Currents. Phys. Rev. Lett. 108, 266602 (2012). https://doi.org/10.1103/PhysRevLett.108.266602

32 D. I. Aladashvili, Z. A. Adamiya, K. G. Lavdovski, E. I. Levin, and B. I. Shklovski, Fiz. Tekh. Poluprovodn. 23, 213 (1989) [Sov. Phys. Semicond. 23, 132 (1989)].

\end{document}